\documentclass[runningheads]{llncs}
\usepackage[T1]{fontenc}

\usepackage{graphicx}
\usepackage{amsmath}
\usepackage{amssymb}
\usepackage{booktabs}
\usepackage{colortbl}
\usepackage{float}
\usepackage{multirow}
\usepackage{array}
\usepackage{placeins}
\usepackage{xcolor}
\usepackage{hyperref}

\usepackage[backend=biber, style=numeric]{biblatex}
\begin{document}
\title{Promptable cancer segmentation using minimal expert-curated data}
\titlerunning{Promptable segmentation using minimal data}

 \author{Lynn Karam \inst{1} \and Yipei Wang \inst{1, 2} \and Veeru Kasivisvanathan \inst{3}, Mirabela Rusu \inst{4}, Yipeng Hu \inst{1, 2} \and Shaheer U. Saeed * \inst{1, 2}}
\authorrunning{Karam et al.}
\institute{Department of Medical Physics and Biomedical Engineering, University College London, London, UK\and
 UCL Hawkes Institute, University College London, London, UK\and
 Division of Surgery and Interventional Science, University College London, London, UK\and
  Department of Radiology; and Department of Urology, Stanford University, California, USA\\
 * Email: \email{shaheer.saeed.17@ucl.ac.uk}}

\maketitle
\begin{abstract}
Automated segmentation of cancer on medical images can aid targeted diagnostic and therapeutic procedures. However, its adoption is limited by the high cost of expert annotations required for training and inter-observer variability in datasets. While weakly-supervised methods mitigate some challenges, using binary histology labels for training as opposed to requiring full segmentation, they require large paired datasets of histology and images, which are difficult to curate. Similarly, promptable segmentation aims to allow segmentation with no re-training for new tasks at inference, however, existing models perform poorly on pathological regions, again necessitating large datasets for training. In this work we propose a novel approach for promptable segmentation requiring only 24 fully-segmented images, supplemented by 8 weakly-labelled images, for training. Curating this minimal data to a high standard is relatively feasible and thus issues with the cost and variability of obtaining labels can be mitigated. By leveraging two classifiers, one weakly-supervised and one fully-supervised, our method refines segmentation through a guided search process initiated by a single-point prompt. Our approach outperforms existing promptable segmentation methods, and performs comparably with fully-supervised methods, for the task of prostate cancer segmentation, while using substantially less annotated data (up to 100X less). This enables promptable segmentation with very minimal labelled data, such that the labels can be curated to a very high standard. \\
Code: \url{github.com/lynnkaram/promptable-cancer-segmentation}
\keywords{Cancer  \and Segmentation \and Prompting methods.}
\end{abstract}
\section{Introduction}

Segmentation of pathological features on medical images helps guide targeted surgical procedures. For prostate cancer, MR image localisation can aid diagnostic biopsies by improving needle placement \cite{chen2023image, barry2001prostate, bostonsci, benelli2020role} and enhance targeted treatments like cryoablation and radiotherapy by minimising damage to healthy tissue \cite{chen2023progress, peschel2003surgery, sekhoacha2022prostate, erinjeri2010cryoablation}. Accurate cancer localisation on medical images is thus crucial for improving diagnostic and therapeutic outcomes.

The cost associated with manually obtaining such localisations, both in terms of time and expertise, has thus far hindered a wider adoption \cite{pocius2024weakly, ratwani2024addressing, alloghani2020systematic}. Limited expert time in clinical settings is coupled with high disagreements on such cancer localisations even between experts \cite{czolbe2021segmentation, chalcroft2021development, ahmad2024early}. Consensus or pixel-level majority votes between localisations from different observers may allow more reliable cancer segmentation. However, these consensus-based segmentations come at an even greater cost in terms of expert time. Such challenges have made adoption practically infeasible in various clinical domains, especially in resource-constrained regions \cite{rosenkrantz2021oncologic, ahmad2024early}. 

To mitigate the cost, recent work has proposed automated segmentation. Deep learning methods \cite{yan2022impact, saeed2022image, yi2024t2}, have shown an ability to accurately delineate boundaries of prostate cancer on MR images, achieving Dice scores in the range of 0.25 - 0.35 compared to radiologists doing the same task. However, such methods require large fully-segmented MR datasets for training, where curating high-quality consensus-based annotations is infeasible due to the associated cost. As a result, most current methods are trained with segmentations from a single institute, which may propagate the subjective biases into the final automated system. This may also be one of the reasons why Dice scores for prostate cancer segmentation have consistently remained low, compared to anatomical segmentation. 

Weakly-supervised methods that aim to localise regions-of-interest (ROIs), using only classification labels of binary ROI presence, have recently been proposed for cancer segmentation \cite{saeed2024competing, li2023weakly}. In these methods, segmentation is learnt solely from a weak label of binary classification indicating whether cancer is present in the image or not, without requiring any segmentations of the MR image. These sorts of weak labels can mitigate subjective biases in the pixel-level segmentation process and can also reduce the time and expertise required to annotate such labels. Objective biopsy-based histology labels indicating binary cancer presence can be used as weak labels to learn cancer segmentation on MR images, without requiring any subjective annotations. Since acquiring pixel-level histopathology labels is infeasible in vivo, such weakly-supervised approaches offer a mechanism to utilise objective information to guide clinical tasks. These approaches require large paired datasets of binary histology labels and MR images for training. However, for prostate cancer, even collecting such paired weak labels proves challenging due to a lack of standardisation in biopsy protocols, reporting, and pairing with imaging data \cite{ahmed2017diagnostic}. 

Recently, promptable methods have emerged as mechanisms to reduce the cost associated with segmentation. Promptable segmentation allows users to specify prompts such as points \cite{kirillov2023segment}, bounding boxes \cite{ma2024segment}, or other sparse annotations \cite{li2024prism, wong2024scribbleprompt}, which can then allow a full segmentation of the ROI. Most of these methods are trained across a wide variety ROIs and are intended to be usable without requiring any new data even from novel classes. However, due to the lack of pathological ROIs on medical images during training, performance remains low for ROIs such as cancer on prostate MR images (consistent with our experiments). Fine-tuning with data for pathological ROIs, such as cancer, may help to improve performance. However, this fine-tuning requires substantial data, almost equivalent to the amount of data required by conventional deep learning methods (as demonstrated in our experiments). Thus these methods suffer from the same challenges as conventional deep learning, which stem from the cost of obtaining high-quality fully-segmented pathology ROIs on medical images. Our experimental results also demonstrate that despite fine-tuning these methods cannot improve performance in challenging tasks such as cancer segmentation.

In this work, we propose a promptable segmentation that can be trained using very few full segmentations (24 segmented images) supplemented by a very small set of weakly-labelled sampled (8 pairs of weak labels and images). In expertise- and data-constrained domains such as cancer segmentation, the use of these few labels during training means that the labels can be feasibly curated to a high standard e.g., following strict protocols for acquiring weak labels based on histology and curating labels through consensus for the full segmentations. Our approach is more data-efficient compared not only to recent state-of-the-art fully-supervised methods \cite{yi2024t2, yan2022impact, saeed2022image} but also to other mixed supervision methods \cite{behzadi2024weakly, rajagopal2024mixed}. The minimal user interaction at inference distinguishes our method from other fully-automated methods, where the user interaction of selecting a point prompt allows our method to achieve accurate segmentation, while being data-efficient.  

We propose to use two classifiers: 1) weakly-supervised classifier; and 2) fully-supervised classifier, to guide a search for the ROI within the image, such that at the end we can obtain a full segmentation. The weakly-supervised classifier is trained using image and histology-based weak label (indicating binary object presence) pairs.
The fully-supervised classifier is trained using image and full segmentation pairs, where the inputs during training are image crops and the outputs are binary labels indicating ROI presence within each crop. After training, crops can be passed to both classifiers and the raw probability (logits) can indicate the likelihood of the object being present within a crop, similar to weak supervision techniques \cite{saeed2024competing, pocius2024weakly, selvaraju2020grad, zhang2023grad}. 

At inference, we require only a point prompt from the user which then leads to a full segmentation. The initial point prompt serves as the location for an initial image crop, which is passed to both the weakly-supervised and fully-supervised classifiers. The scores from the classifier are combined and thresholded to determine whether the crop should be marked as ROI presence positive or negative. The crop is then moved in a set pattern (explained in the methods), determining ROI presence at each location. The ROI-positive crops are then combined into a full segmentation. This approach allows a promptable segmentation to be trained using very limited data, such that this data can be curated to a very high standard.

The contributions of our work are summarised:
\begin{enumerate}
\item proposing a novel framework for promptable segmentation, which can be trained with minimal labels
\item evaluating the approach for a clinically challenging task of prostate cancer segmentation using a large real world clinical dataset
\item demonstrating superior performance compared to other promptable methods and comparable performance to fully-supervised learning which requires 100X the amount of data for training
\item presenting an open-source implementation of our approach at: \url{github.com/lynnkaram/promptable-cancer-segmentation}
\end{enumerate}

\section{Methods}

\subsection{Weakly-supervised classifier}

The weakly-supervised classifier $f(\cdot;  \theta): \mathcal{X}\rightarrow [0,1]$ takes an input image $x \in \mathcal{X}$ and outputs a probability of ROI presence in the range $[0,1]$, where $\mathcal{X}$ represents the domain of input image samples and $\theta$ denotes the weights of the neural network.

The classifier is trained on a dataset consisting of pairs of images and binary labels $\{(x_{i}, y_{i})\}_{i=1}^{N}$, where $y_{i} \in \{0, 1\}$ indicating whether the ROI is present in the image or not.

The binary cross-entropy loss is used for training the classifier:
\begin{align}
\mathcal{L}_f = - \frac{1}{N} \sum_{i=1}^{N} \left[ y_{i} \log( f(x_i; \theta)) + (1 - y_{i}) \log (1 - f(x_i; \theta)) \right]
\end{align}

Where the optimal weights for the neural network are given by:

\begin{align}
    \theta^* = \arg\min_\theta \mathcal{L}_f
\end{align}

\noindent\textbf{Image crops at inference:}
A crop from the image $x_i$ is defined as $x_{i, c}^{(w, h, d)}\in\mathcal{X}_c$, where $(w, h, d)$ denotes the locations of a fixed-size crop within the image, in the width, height and depth dimensions respectively, and $\mathcal{X}_c$ denotes the domain of crops. The subscript $c$ is used to indicate the number of crops in a set of crops $\{x_{i, c}^{(w, h, d)}\}_{c=1}^C$, where $C$ is the number of crops per image.

The score for the image crop can then be computed using $f(x_{i, c}^{(w, h, d)}; \theta^*) \in [0, 1]$. This strategy of training with full images and inference on crops or portions is similar to recent weak-supervision literature \cite{saeed2024competing, pocius2024weakly, zhang2023grad, selvaraju2020grad}.

\subsection{Fully-supervised classifier}

The fully-supervised classifier $g(\cdot; \phi): \mathcal{X}_c\rightarrow [0,1]$ takes a crop as input $x_{i, c}^{(w, h, d)}\in\mathcal{X}_c$ and outputs a probability of ROI presence in the range $[0,1]$, where $\phi$ denotes the neural network weights. 

The classifier is trained on a dataset consisting of pairs of crops and binary labels 
$\{\{(x_{i, c}^{(w,h,d)}, ~z_{i, c}^{(w,h,d))} \}_{c=1}^{C}\}_{i=1}^{M}$, where $C$ is the number of crops per image and there are a total of $M$ images, and $z_{i, c}^{(w,h,d))}\in\{0,1\}$ indicates whether an ROI is present within the crop or not. Although a full segmentation can be used to construct such labels, other mechanisms may be used depending on data availability e.g., more dense tracked biopsies may be able to indicate crop-level presence in an objective manner as well.

The binary cross-entropy loss is used for training the classifier:
\begin{align}
\mathcal{L}_g = - \frac{1}{M*C} \sum_{i=1}^{M} \sum_{c=1}^{C} \left[ z_{i, c}^{(w,h,d))} \log( g(x_{i, c}^{(w, h, d)}; \phi) + (1 - z_{i, c}^{(w,h,d))}) \log (1 - g(x_{i, c}^{(w, h, d)}; \phi)) \right]
\end{align}

Where the optimal weights for the neural network are given by:

\begin{align}
    \phi^* = \arg\min_\phi \mathcal{L}_g
\end{align}

\noindent\textbf{Image crops at inference:}
The score for the image crop at inference can be computed using $g(x_{i, c}^{(w, h, d)}; \phi^*)\in[0,1]$.

\begin{figure}
    \centering
    \includegraphics[width=0.8\textwidth]{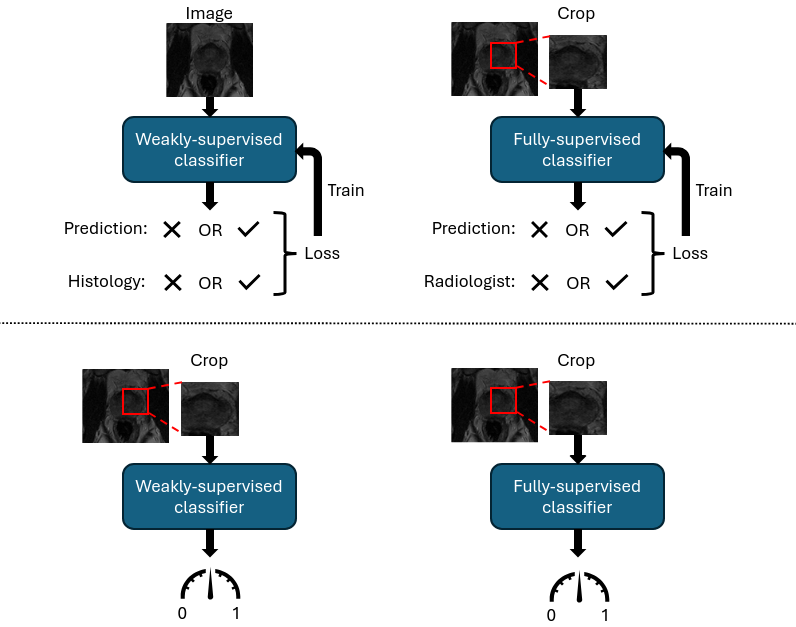}
    \caption{The training (top) and inference (bottom) pipelines for the weakly-supervised and fully-supervised classifiers.}
    \label{fig:classifiers}
\end{figure}

\subsection{Point prompt and search}

\noindent\textbf{Point prompt:} During promptable segmentation, the user provides a point prompt at coordinates $(w_0, h_0, d_0)$. The crop at these coordinates is then $x_{i, c}^{(w_0,h_0,d_0)}$. The subscripts $0$ denote the initial time-step $t=0$, which iterates as the crop moves.

\noindent\textbf{Crop scoring:} We define a joint score as follows (where $\alpha\in[0,1]$ is a hyper-parameter): 

\begin{align}
    S = \alpha\times f(x_{i, c}^{(w_t, h_t, d_t)}; \theta^*) ~ + ~(1-\alpha) \times g(x_{i, c}^{(w_t, h_t, d_t)}; \phi^*)
\end{align}

The score is thresholded using a threshold $\tau$:

\begin{align}
S_{\tau, t} = 
\begin{cases} 
1, & \text{if } S > \tau, \\
0, & \text{otherwise}.
\end{cases}
\end{align}

Here, $S_{\tau, t}$ indicates whether a crop at step $t$ is marked as ROI positive or negative.

\noindent\textbf{Crop movement:}
The crop then moves following a spline, which is a spiral in our work. The spiral strategy is more computationally efficient compared to a full sliding window approach due to a lower number of crops processed and roughly follows the spherical shape of the prostate gland. In preliminary results, this strategy showed performance improvements compared to strategies such as random search. As a crop moves, $S_t$ is computed for each crop. The movement is defined as a spiral spline with parameters $s$ and $\mu$, denoting scale and total number of steps for a full circle, respectively. Steps in the spline are denoted by $t$, where the radius for any given step is computed as $r = t/s$ and the angle is given by $\beta=2\times \pi \times (t/ \mu)$. The crop movement is defined by:

$$\delta w = r \times cos (\beta)$$
\begin{align}
    \delta  h = r \times sin(\beta)
\end{align}
$$\delta  d = 0$$

The new crop location can then be given by:

\begin{align}
(w_{t+1}, h_{t+1}, d_{t+1}) = (w_{t} + \delta w, h_{t} + \delta h, d_{t} + \delta d)    
\end{align}

\noindent\textbf{Final segmentation:} As $t$ iterates until a fixed number of steps $T$, the joint scores for each crop are collected $\{S_{\tau, t}\}_{t=0}^T$. Each location where $S_{\tau, t} = 1$, is combined to form the full segmentation. In practice the spiral search may be performed $n$ times with varying parameters, and then conducting a pixel-level majority vote, such that a pixel-level segmentation map can be obtained. 

\begin{figure}[!ht] 
    \centering
    \includegraphics[width=0.98
    \textwidth]{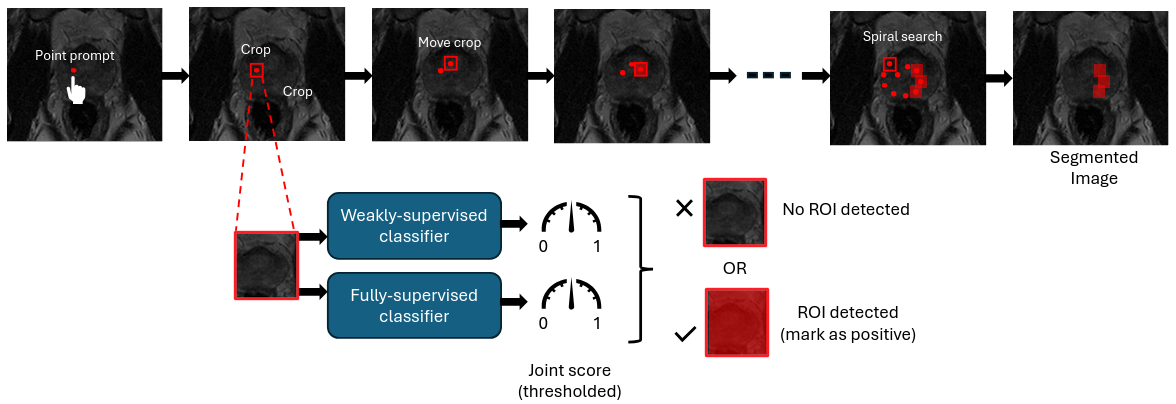} 
    \caption{The point prompt leads to a spiral search where each crop is scored by the two classifiers to determine whether to mark a crop as positive or negative. Combining all positive crops gives the final segmentation.}
    \label{fig:method}
\end{figure}

The framework is summarised in Fig. \ref{fig:method}.

\section{Experiments}

\subsection{Dataset}

The dataset used in this work consists of 500 MR images from men over the age of 30, suspected of having prostate cancer. The images were bi-parametric pelvic MR scans with channels corresponding to T2-weighted and diffusion-weighted images. These images were collected as part of the [anonymous] trials. Each image had a binary histology label indicating cancer presence and a consensus-based radiologist annotation of the full segmentation of cancer.

Out of these images, 200 were reserved as the holdout set and the rest were used for development (including training other methods for comparison).

\subsection{Network architectures}

Both classifiers followed a 3D convolutional neural networks (CNN) architecture \cite{o2015introduction, krizhevsky2012imagenet}. The classifiers used 16, 32, and 64 filters in three convolutional layers, each using ReLU activation. Each convolutional layer was followed by a MaxPooling3D layer (2,2,2) to reduce spatial dimensions. The extracted features were then flattened and passed through two fully connected (dense) layers. The model was optimised using Adam. For other hyper-parameter settings please refer to experiments (specified as appropriate) or code available in open-source repository.

The training time for our method was 48 hours on a single Nvidia Tesla V100 GPU.

\subsection{Experimental protocols}

\subsubsection{Comparisons:}

We compared our approach to a fully-supervised U-Net \cite{ronneberger2015u}, a fine-tuned promptable SAM \cite{wang2024promise} and MedSAM \cite{ma2024segment}, and recent state-of-the-art methods in prostate cancer segmentation \cite{yi2024t2, yan2022impact}. The U-Net \cite{ronneberger2015u} was pre-trained using fully-supervised learning with 1000 samples from the PI-CAI (Prostate Imaging: Cancer AI) challenge data \cite{saha2022pi} and then fine-tuned using 200 samples from our dataset. This, along with other fully-supervised state-of-the-art methods using similar architectures \cite{yi2024t2, yan2022impact}, serve as upper-bound reference performance in the task of prostate cancer segmentation. SAM \cite{wang2024promise} and MedSAM \cite{ma2024segment} were fine-tuned for our dataset using 200 samples, and represent promptable segmentation methods. Note that all methods that we compared with used approximately 10X-100X the amount of data that is used for training our promptable segmentation approach.

\subsubsection{Ablations:}

To investigate the impact of hyper-parameters and the amount of training data for our classifiers, we conducted ablation studies. The ablation studies investigated crop sizes, amount of training data for classifiers and the impact of the threshold $\tau$. We also investigated variance in our method with respect to the initial point prompt location within the cancerous area.

\section{Results}

\subsection{Comparisons}

Tab. \ref{tab:comparison} shows that our method outperforms other promptable methods while using 100X less data for training. Compared to both SAM and MedSAM, the differences were significant (paired t-test with significance level 0.05, p-values 0.0004 and 0.0032). For the comparison with the U-Net model, which is an upper-bound performance, statistical significance was not found for the comparison (p-value 0.07). Statistical tests were not conducted for the other two methods as their test data were different from our dataset, however, the differences to these methods were small (only 0.031 Dice difference to Yi et al \cite{yi2024t2}).

\begin{table}[!ht]
    \centering
        \caption{Comparison to recent state-of-the-art methods for both promptable and non-promptable methods.}
    \begin{tabular}{|c|c|c|c|}
        \hline
                     &Training Data & Dice  &  Variance w.r.t. prompt\\ \hline
        Ours         & 8 weak + 24 full               & 0.3085 $\pm$ 0.1405 & $\pm$ 0.2163 \\ \hline
        SAM          & 200 full               & 0.2361 $\pm$ 0.1074& $\pm$ 0.2359\\ \hline
        SAM          & 32 full               & 0.1587 $\pm$ 0.1319& $\pm$ 0.1729\\ \hline

        MedSAM       & 200 full               & 0.2673 $\pm$ 0.1384 & $\pm$ 0.2103\\ \hline
        MedSAM          & 32 full               & 0.1846 $\pm$ 0.1492& $\pm$ 0.1837\\ \hline

        U-Net        & 1200 full               & 0.3275 $\pm$ 0.1783 & -\\ \hline
        U-Net          & 32 full               & 0.1923 $\pm$ 0.1819& -\\ \hline

        Yi et al \cite{yi2024t2} & 275 full & 0.3392 $\pm$ 0.2358& -\\ \hline
        Yan et al \cite{yan2022impact} & 503 full & 0.3300 $\pm$ 0.1800& -\\
        \hline
    \end{tabular}
    \label{tab:comparison}
\end{table}

\subsection{Ablations}

\subsubsection{Impact of amount of training data:}

Tab. \ref{tab:classifier_performance} presented the classifier and segmentation performances when the classifiers were trained with varying amounts of data. Convergence for the weakly-supervised classifier was observed with 8 samples and for the fully-supervised classifier with 24 samples. This low number of samples may stem from the fact that classification tasks are easier to learn due to their low-dimensional solution space compared to tasks like segmentation, which require more samples for learning. This is also in-line with recent works \cite{saeed2024competing}.

\begin{table}[!ht]
    \centering
    \caption{Performance of classifier and segmentation based on training sample size. WSC is the weakly supervised classifier and FSC is the fully-supervised classifier.}
{ 
    \begin{tabular}{|c|c|c|c|c|}
        \hline
        {WSC Samples} & {FSC Samples} & WSC Acc. & FSC Acc. & {Dice}  \\
        \hline
        4 & 24 & 0.782 & 0.582 & 0.211 \\
        8 & 24 & 0.782 & 0.723 & 0.305 \\
        12 & 24 & 0.782 & 0.716 & 0.298 \\
        16 & 24 & 0.782 & 0.732 & 0.307 \\
        \hline
        8 &  8 & 0.643 & 0.723 & 0.204 \\
        8 & 16 & 0.715 & 0.723 & 0.261\\
        8 & 24 & 0.782 & 0.723 & 0.305 \\
        8 & 32 & 0.896 & 0.723 & 0.309 \\
        8 & 40 & 0.899 & 0.723 & 0.308 \\
        \hline
    \end{tabular}
    }
    \label{tab:classifier_performance}
\end{table}

\subsubsection{Impact of crop sizes:}

\begin{figure}[H] 
    \centering
    \includegraphics[width=0.8\textwidth]{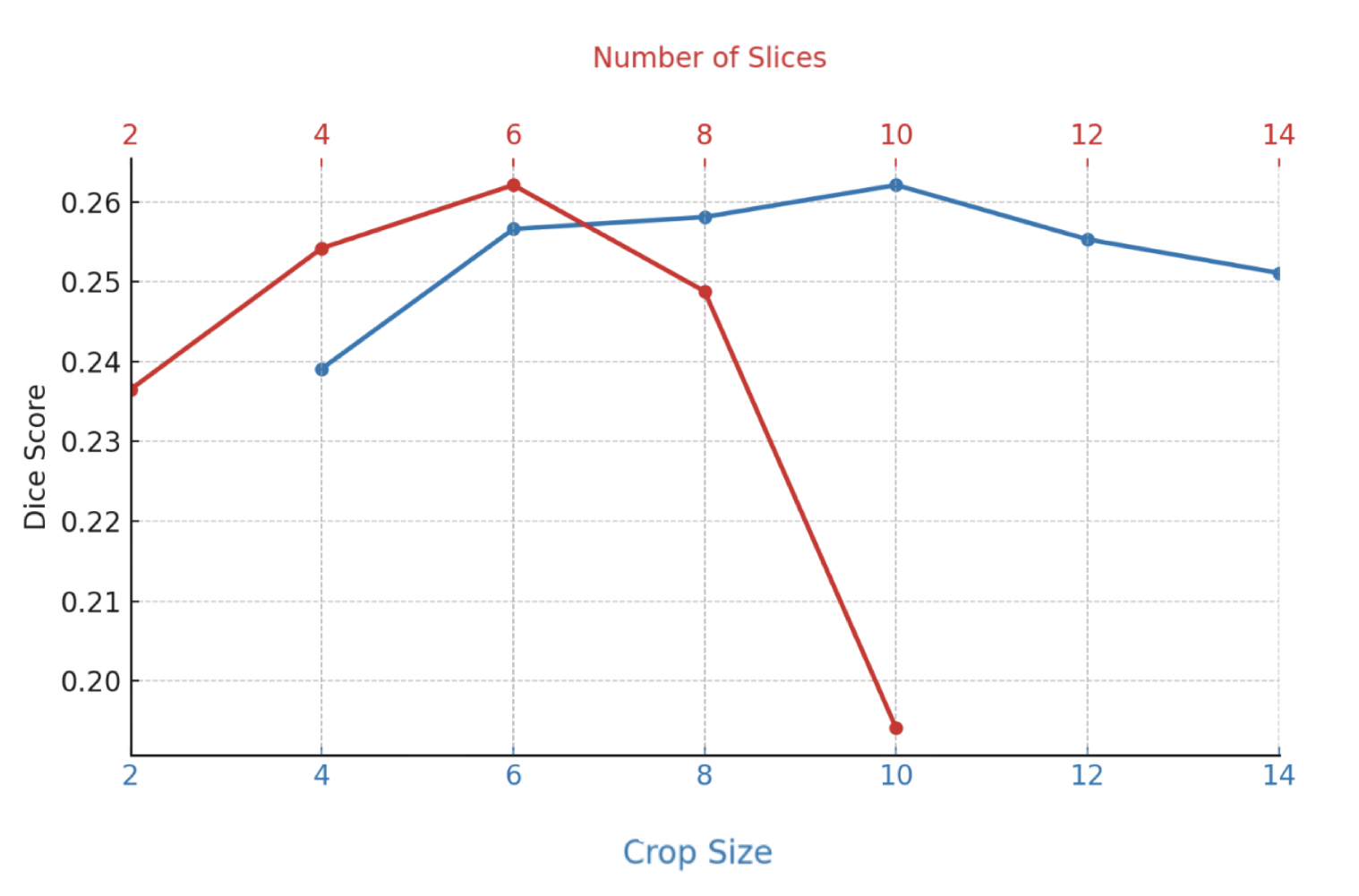} 
        \caption{Plot of crop size $(w, h, d)$ against dice score. $(w, h)$ on the bottom axis and $d$ on the top axis.}
    \label{fig:res_crop}
\end{figure}

As shown in Fig. \ref{fig:res_crop}, crop sizes substantially impact performance. A crop size of 10 pixels in the height and width dimensions (i.e. the x- and y-directions) yielded the highest dice score of $0.2621 \pm 0.2106$. And a crop size of 6 in the depth dimension, which corresponds to number of slices, yielded the highest dice score of $0.2621 \pm 0.2106$. The optimal crop size was thus selected as 10x10x6.

\subsubsection{Impact of search strategies}

As shown in Tab. \ref{tab:my_label}, the spiral search strategy is both time-efficient and approaches human-level performance, compared to other common strategies.

\begin{table}[!ht]
    \centering
    \caption{The time and Dice scores for different search strategies compared.}
    \begin{tabular}{|c|c|c|}
    \hline
    Strategy & Time & Dice\\
    Spiral (Ours) &  4.2s & 0.2967\\
    Sliding Window   & 12.8s & 0.2957\\
    Random Search & 5.9s & 0.2431\\
    Expert Human & 439.6s & 0.2971\\
\hline
    \end{tabular}
    \label{tab:my_label}
\end{table}

\subsubsection{Impact of hyper-parameters}

As summarised in Tab. \ref{table:thresholds}, the optimal value for $\tau$ was 0.05, which yielded the highest Dice. Only the difference between 0.01 to 0.05 was statistically significant (p-value 0.0312), and we did not observe statistical significance for the comparison between 0.05 and 0.10 (p-value 0.1130).

Similarly, Tab. \ref{table:thresholds} also presents the impact of other hyper-parameters, including the number of steps $T$, total number of steps for a full circle in the spiral $\mu$, the parameter controlling the balance between weak and full supervision $\alpha$ and the number of times spiral search is repeated for a single sample $n$.

\begin{table}[!ht]
    \centering
    \caption{Dice scores and standard deviation for different hyper-parameters (statistically significant performance improvement emboldened; where none or more than one emboldened, statistical significance was not found for the comparison).}
    \begin{tabular}{|c|c|}
        \hline
        { $\tau$} & {Dice} \\
        \hline
        0.01  & 0.2286 $\pm$ 0.1785 \\
        0.05  & \textbf{0.2689 $\pm$ 0.2163} \\
        0.10   & \textbf{0.2598 $\pm$ 0.2389} \\
        \hline
        & \\
        \hline
        $T$ & \\
        \hline
        60 & 0.2579$\pm$0.1983\\
        80 & 0.2611$\pm$0.2189\\
        100 & 0.2573$\pm$0.2055\\
        \hline
        & \\
        \hline
        $\mu$ & \\
        \hline
        200 & 0.2614$\pm$0.2018\\
        400 & 0.2571$\pm$0.2023\\
        600 & 0.2510$\pm$0.1997\\
        \hline
        & \\
        \hline
        $\alpha$ & \\
        \hline
        0.25 & \textbf{0.2753$\pm$0.1932}\\
        0.50 & \textbf{0.2749$\pm$0.2140}\\
        0.75 & 0.2521$\pm$0.2276\\
        \hline
        & \\
        \hline
        $n$ & \\
        \hline
        4 & 0.2732$\pm$0.2113\\
        6 & 0.2873$\pm$0.1803\\
        8 & 0.2838$\pm$0.2001\\
        \hline
    \end{tabular}
    \label{table:thresholds}
\end{table}

\subsection{Qualitative analysis}

\begin{figure}[!ht] 
    \centering
    \includegraphics[width=1\textwidth]{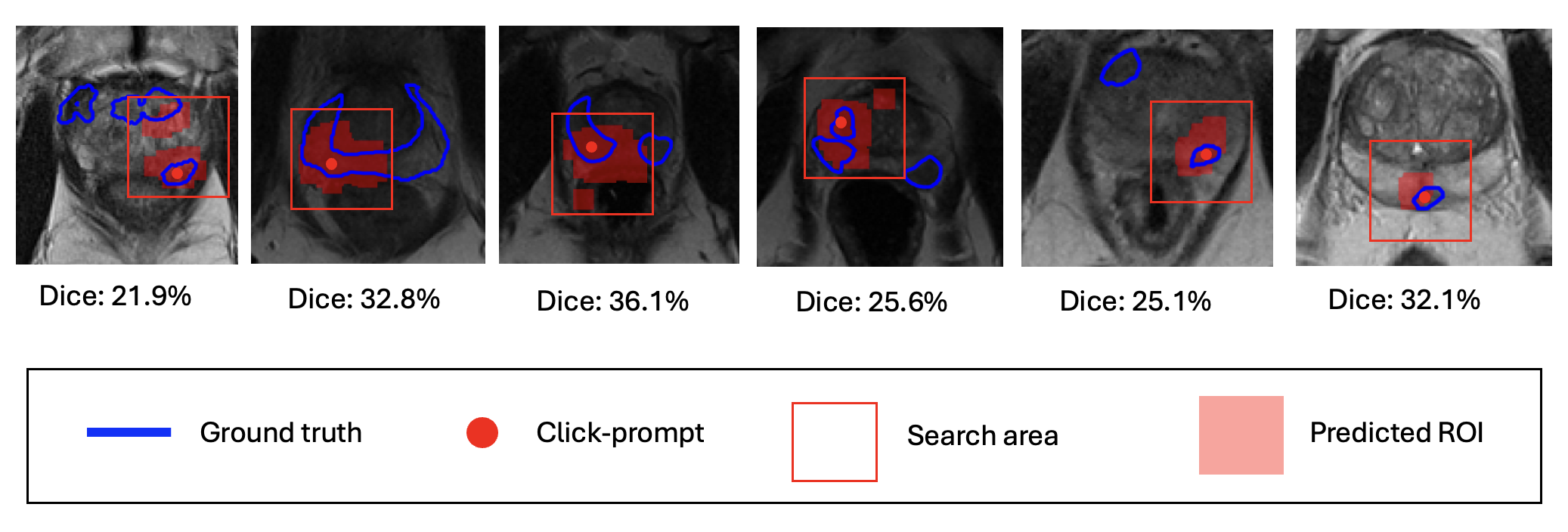} 
    \caption{Samples of MR scans. Ground truth is in blue, predicted is in red.}
    \label{fig:samples}
\end{figure}

Fig. \ref{fig:samples} shows examples of the lesions predicted after prompting. Note that we only search a specified area in the spiral search and thus final scores were computed across different click locations to cover the entire cancerous lesion. As seen in the randomly picked examples, our method shows substantial agreement with a radiologist performing the same task.\\

\section{Discussion}

Our results show that the proposed framework allows effective prostate cancer localisation on MR images from only a point prompt. Our approach outperforms other common promptable segmentation methods while performing comparably to fully-supervised non-promptable methods, all while using substantially less data. For training, our framework only requires 8 binary histology labels indicating cancer presence, paired with MR images, and 24 radiologist segmentations of cancer. This minimal data requirement means that it is feasible to curate the data to a high standard e.g., by following strict reporting protocols for histology or curating the radiologist annotations through consensus. In other methods that we compared against, the training requires hundreds or thousands of radiologist annotated samples, which makes it infeasible to curate labels through consensus or majority votes. This means that as other methods use single observer labels for training, the subjective biases of the observers may be passed into the final trained cancer segmentation system. In contrast our method uses objective histology labels coupled with a very small set of radiologist annotations, where it is feasible to ensure that the annotations minimise the subjective biases through mechanisms such as consensus. 

The over prediction is likely caused by over-representation of cancer positive cases in the data. Tuning the termination threshold $\tau$ holds potential to balance over- and under-prediction. Over-prediction may be favourable for biopsy sampling, to represent not only the most significant cancer but also surrounding areas to determine spread. Future work could explore more detailed analysis of balancing over- and under-prediction for other applications. 

Our method has potential to enable promptable segmentation in a variety of domains where data is scarce or where annotations from experts are expensive. It should be noted that while the classifiers used in our framework converged with minimal examples, there may be applications where more data is required. And while the spiral search was effective for prostate cancer segmentation, possibly due to the roughly spherical shape of the gland, other strategies may need to be explored for other applications.

\section{Conclusion}

In this work we proposed a framework to enable promptable segmentation of complex structures such as cancer in the prostate gland, using very minimal data for training. Evaluating the approach with data from real prostate cancer patients, we found performance comparable to state-of-the-art methods while requiring a fraction of the annotated data for training. Our approach has potential to reduce the segmentation burden in a variety of applications where data or annotations are scarce including cancer localisation in other parts of the body, rare disease localisation and anomaly detection.

\section*{Acknowledgements}

This work is supported by the International Alliance for Cancer Early Detection, an alliance between Cancer Research UK [EDDAPA-2024/100014] \& [C73666/A31378], Canary Center at Stanford University, the University of Cambridge, OHSU Knight Cancer Institute, University College London and the University of Manchester; and National Institute for Health Research University College London Hospitals Biomedical Research Centre.

\printbibliography

@article{saeed2024competing,
  title={Competing for pixels: a self-play algorithm for weakly-supervised semantic segmentation},
  author={Saeed, Shaheer U and Huang, Shiqi and Ramalhinho, Jo{\~a}o and Gayo, Iani JMB and Monta{\~n}a-Brown, Nina and Bonmati, Ester and Pereira, Stephen P and Davidson, Brian and Barratt, Dean C and Clarkson, Matthew J and others},
  journal={IEEE Transactions on Pattern Analysis and Machine Intelligence},
  year={2024},
  publisher={IEEE}
}

@inproceedings{pocius2024weakly,
  title={Weakly supervised localisation of prostate cancer using reinforcement learning for bi-parametric MR images},
  author={Pocius, Martynas and Yan, Wen and Barratt, Dean C and Emberton, Mark and Clarkson, Matthew J and Hu, Yipeng and Saeed, Shaheer U},
  booktitle={2024 IEEE International Symposium on Biomedical Imaging (ISBI)},
  pages={1--5},
  year={2024},
  organization={IEEE}
}

@article{rosenkrantz2021oncologic,
  title={Oncologic errors in diagnostic radiology: a 10-year analysis based on medical malpractice claims},
  author={Rosenkrantz, Andrew B and Siegal, Dana and Skillings, Jillian A and Muellner, Ada and Nass, Sharyl J and Hricak, Hedvig},
  journal={Journal of the American College of Radiology},
  volume={18},
  number={9},
  pages={1310--1316},
  year={2021},
  publisher={Elsevier}
}

@article{ratwani2024addressing,
  title={Addressing AI algorithmic bias in health care},
  author={Ratwani, Raj M and Sutton, Karey and Galarraga, Jessica E},
  journal={JAMA},
  volume={332},
  number={13},
  pages={1051--1052},
  year={2024},
  publisher={American Medical Association}
}

@article{ahmad2024early,
  title={Early cancer detection using deep learning and medical imaging: A survey},
  author={Ahmad, Istiak and Alqurashi, Fahad},
  journal={Critical Reviews in Oncology/Hematology},
  pages={104528},
  year={2024},
  publisher={Elsevier}
}

@article{alloghani2020systematic,
  title={A systematic review on supervised and unsupervised machine learning algorithms for data science},
  author={Alloghani, Mohamed and Al-Jumeily, Dhiya and Mustafina, Jamila and Hussain, Abir and Aljaaf, Ahmed J},
  journal={Supervised and unsupervised learning for data science},
  pages={3--21},
  year={2020},
  publisher={Springer}
}

@article{chen2023progress,
  title={Progress in the cryoablation and cryoimmunotherapy for tumor},
  author={Chen, Zenan and Meng, Liangliang and Zhang, Jing and Zhang, Xiao},
  journal={Frontiers in immunology},
  volume={14},
  pages={1094009},
  year={2023},
  publisher={Frontiers Media SA}
}

@article{erinjeri2010cryoablation,
  title={Cryoablation: mechanism of action and devices},
  author={Erinjeri, Joseph P and Clark, Timothy WI},
  journal={Journal of Vascular and Interventional Radiology},
  volume={21},
  number={8},
  pages={S187--S191},
  year={2010},
  publisher={Elsevier}
}

@article{sekhoacha2022prostate,
  title={Prostate cancer review: genetics, diagnosis, treatment options, and alternative approaches},
  author={Sekhoacha, Mamello and Riet, Keamogetswe and Motloung, Paballo and Gumenku, Lemohang and Adegoke, Ayodeji and Mashele, Samson},
  journal={Molecules},
  volume={27},
  number={17},
  pages={5730},
  year={2022},
  publisher={MDPI}
}

@article{peschel2003surgery,
  title={Surgery, brachytherapy, and external-beam radiotherapy for early prostate cancer},
  author={Peschel, Richard E and Colberg, John W},
  journal={The lancet oncology},
  volume={4},
  number={4},
  pages={233--241},
  year={2003},
  publisher={Elsevier}
}

@article{barry2001prostate,
  title={Prostate-specific--antigen testing for early diagnosis of prostate cancer},
  author={Barry, Michael J},
  journal={New England Journal of Medicine},
  volume={344},
  number={18},
  pages={1373--1377},
  year={2001},
  publisher={Mass Medical Soc}
}

@article{benelli2020role,
  title={The role of MRI/TRUS fusion biopsy in the diagnosis of clinically significant prostate cancer},
  author={Benelli, Andrea and Vaccaro, Chiara and Guzzo, Sonia and Nedbal, Carlotta and Varca, Virginia and Gregori, Andrea},
  journal={Therapeutic Advances in Urology},
  volume={12},
  pages={1756287220916613},
  year={2020},
  publisher={SAGE Publications Sage UK: London, England}
}

@article{chen2023image,
  title={Image registration: Fundamentals and recent advances based on deep learning},
  author={Chen, Min and Tustison, Nicholas J and Jena, Rohit and Gee, James C},
  journal={Machine Learning for Brain Disorders},
  pages={435--458},
  year={2023},
  publisher={Springer}
}

@article{selvaraju2020grad,
  title={Grad-CAM: visual explanations from deep networks via gradient-based localization},
  author={Selvaraju, Ramprasaath R and Cogswell, Michael and Das, Abhishek and Vedantam, Ramakrishna and Parikh, Devi and Batra, Dhruv},
  journal={International journal of computer vision},
  volume={128},
  pages={336--359},
  year={2020},
  publisher={Springer}
}

@article{zhang2023grad,
  title={Grad-CAM-based explainable artificial intelligence related to medical text processing},
  author={Zhang, Hongjian and Ogasawara, Katsuhiko},
  journal={Bioengineering},
  volume={10},
  number={9},
  pages={1070},
  year={2023},
  publisher={MDPI}
}

@inproceedings{li2024prism,
  title={Prism: A promptable and robust interactive segmentation model with visual prompts},
  author={Li, Hao and Liu, Han and Hu, Dewei and Wang, Jiacheng and Oguz, Ipek},
  booktitle={International Conference on Medical Image Computing and Computer-Assisted Intervention},
  pages={389--399},
  year={2024},
  organization={Springer}
}

@article{wang2024promise,
  title={ProMISe: Promptable Medical Image Segmentation using SAM},
  author={Wang, Jinfeng and Song, Sifan and Wang, Xinkun and Wang, Yiyi and Miao, Yiyi and Su, Jionglong and Zhou, S Kevin},
  journal={arXiv preprint arXiv:2403.04164},
  year={2024}
}

@inproceedings{kirillov2023segment,
  title={Segment anything},
  author={Kirillov, Alexander and Mintun, Eric and Ravi, Nikhila and Mao, Hanzi and Rolland, Chloe and Gustafson, Laura and Xiao, Tete and Whitehead, Spencer and Berg, Alexander C and Lo, Wan-Yen and others},
  booktitle={Proceedings of the IEEE/CVF international conference on computer vision},
  pages={4015--4026},
  year={2023}
}

@article{ma2024segment,
  title={Segment anything in medical images and videos: Benchmark and deployment},
  author={Ma, Jun and Kim, Sumin and Li, Feifei and Baharoon, Mohammed and Asakereh, Reza and Lyu, Hongwei and Wang, Bo},
  journal={arXiv preprint arXiv:2408.03322},
  year={2024}
}

@inproceedings{wong2024scribbleprompt,
  title={Scribbleprompt: fast and flexible interactive segmentation for any biomedical image},
  author={Wong, Hallee E and Rakic, Marianne and Guttag, John and Dalca, Adrian V},
  booktitle={European Conference on Computer Vision},
  pages={207--229},
  year={2024},
  organization={Springer}
}

@inproceedings{ronneberger2015u,
  title={U-net: Convolutional networks for biomedical image segmentation},
  author={Ronneberger, Olaf and Fischer, Philipp and Brox, Thomas},
  booktitle={Medical image computing and computer-assisted intervention--MICCAI 2015: 18th international conference, Munich, Germany, October 5-9, 2015, proceedings, part III 18},
  pages={234--241},
  year={2015},
  organization={Springer}
}

@article{krizhevsky2012imagenet,
  title={Imagenet classification with deep convolutional neural networks},
  author={Krizhevsky, Alex and Sutskever, Ilya and Hinton, Geoffrey E},
  journal={Advances in neural information processing systems},
  volume={25},
  year={2012}
}

@article{o2015introduction,
  title={An introduction to convolutional neural networks},
  author={O'shea, Keiron and Nash, Ryan},
  journal={arXiv preprint arXiv:1511.08458},
  year={2015}
}

@inproceedings{czolbe2021segmentation,
  title={Is segmentation uncertainty useful?},
  author={Czolbe, Steffen and Arnavaz, Kasra and Krause, Oswin and Feragen, Aasa},
  booktitle={Information Processing in Medical Imaging: 27th International Conference, IPMI 2021, Virtual Event, June 28--June 30, 2021, Proceedings 27},
  pages={715--726},
  year={2021},
  organization={Springer}
}

@inproceedings{chalcroft2021development,
  title={Development and evaluation of intraoperative ultrasound segmentation with negative image frames and multiple observer labels},
  author={Chalcroft, Liam F and Qu, Jiongqi and Martin, Sophie A and Gayo, Iani JMB and Minore, Giulio V and Singh, Imraj RD and Saeed, Shaheer U and Yang, Qianye and Baum, Zachary MC and Altmann, Andre and others},
  booktitle={Simplifying Medical Ultrasound: Second International Workshop, ASMUS 2021, Held in Conjunction with MICCAI 2021, Strasbourg, France, September 27, 2021, Proceedings 2},
  pages={25--34},
  year={2021},
  organization={Springer}
}

@article{bostonsci,
    author = {Boston Scientific} ,
    title = {Cryoablation Treatment Guide},
    journal = {Advancing science for life} ,
    year = 2021
}

@article{saha2022pi,
  title={The PI-CAI challenge: public training and development dataset},
  author={Saha, Anindo and Twilt, Jasper Jonathan and Bosma, Joeran Sander and van Ginneken, Bram and Yakar, Derya and Elschot, Mattijs and Veltman, Jeroen and F{\"u}tterer, Jurgen and de Rooij, Maarten and Huisman, Henkjan},
  journal={(No Title)},
  year={2022},
  publisher={Zenodo}
}

@inproceedings{yan2022impact,
  title={The impact of using voxel-level segmentation metrics on evaluating multifocal prostate cancer localisation},
  author={Yan, Wen and Yang, Qianye and Syer, Tom and Min, Zhe and Punwani, Shonit and Emberton, Mark and Barratt, Dean and Chiu, Bernard and Hu, Yipeng},
  booktitle={International Workshop on Applications of Medical AI},
  pages={128--138},
  year={2022},
  organization={Springer}
}

@article{saeed2022image,
  title={Image quality assessment by overlapping task-specific and task-agnostic measures: application to prostate multiparametric MR images for cancer segmentation},
  author={Saeed, Shaheer U and Yan, Wen and Fu, Yunguan and Giganti, Francesco and Yang, Qianye and Baum, Zachary and Rusu, Mirabela and Fan, Richard E and Sonn, Geoffrey A and Emberton, Mark and others},
  journal={Machine Learning in Biomedical Imaging (MELBA)},
  year={2022}
}

@article{yi2024t2,
  title={T2-Only Prostate Cancer Prediction by Meta-Learning from Bi-Parametric MR Imaging},
  author={Yi, Weixi and Wang, Yipei and Thorley, Natasha and Ng, Alexander and Punwani, Shonit and Kasivisvanathan, Veeru and Barratt, Dean C and Saeed, Shaheer Ullah and Hu, Yipeng},
  journal={arXiv preprint arXiv:2411.07416},
  year={2024}
}

@article{li2023weakly,
  title={Weakly supervised histopathology image segmentation with self-attention},
  author={Li, Kailu and Qian, Ziniu and Han, Yingnan and Chang, Eric I-Chao and Wei, Bingzheng and Lai, Maode and Liao, Jing and Fan, Yubo and Xu, Yan},
  journal={Medical Image Analysis},
  volume={86},
  pages={102791},
  year={2023},
  publisher={Elsevier}
}

@article{ahmed2017diagnostic,
  title={Diagnostic accuracy of multi-parametric MRI and TRUS biopsy in prostate cancer (PROMIS): a paired validating confirmatory study},
  author={Ahmed, Hashim U and Bosaily, Ahmed El-Shater and Brown, Louise C and Gabe, Rhian and Kaplan, Richard and Parmar, Mahesh K and Collaco-Moraes, Yolanda and Ward, Katie and Hindley, Richard G and Freeman, Alex and others},
  journal={The Lancet},
  volume={389},
  number={10071},
  pages={815--822},
  year={2017},
  publisher={Elsevier}
}

@article{behzadi2024weakly,
  title={Weakly-supervised deep learning model for prostate cancer diagnosis and gleason grading of histopathology images},
  author={Behzadi, Mohammad Mahdi and Madani, Mohammad and Wang, Hanzhang and Bai, Jun and Bhardwaj, Ankit and Tarakanova, Anna and Yamase, Harold and Nam, Ga Hie and Nabavi, Sheida},
  journal={Biomedical Signal Processing and Control},
  volume={95},
  pages={106351},
  year={2024},
  publisher={Elsevier}
}

@article{rajagopal2024mixed,
  title={Mixed supervision of histopathology improves prostate cancer classification from MRI},
  author={Rajagopal, Abhejit and Westphalen, Antonio C and Velarde, Nathan and Simko, Jeffry P and Nguyen, Hao and Hope, Thomas A and Larson, Peder EZ and Magudia, Kirti},
  journal={IEEE Transactions on Medical Imaging},
  year={2024},
  publisher={IEEE}
}

\end{document}